\documentclass[letter]{aa}  

\usepackage{natbib}
\usepackage{graphicx}
\usepackage{txfonts}
\usepackage{hyperref}

\usepackage{amstext}

\begin{document} 

  \title{Long uninterrupted photometric observations of the Wolf-Rayet star EZ CMa by the Toronto {\em{BRITE}} satellite reveal a very fast apsidal motion}
  
\titlerunning{Fast apsidal motion in the binary orbit of EZ CMa}
  \author{W. Schmutz\inst{1}
     \and
           G. Koenigsberger\inst{2}
          }

  \institute{Physikalisch-Meteorologisches Observatorium Davos and World Radiation Center,
              Dorfstrasse 33, CH-7260 Davos Dorf, Switzerland\\
              \email{werner.schmutz@pmodwrc.ch}
         \and
             Instituto de Ciencias F\'isicas, Universidad Nacional Aut\'onoma de M\'exico, Ave. Universidad S/N, Cuernavaca, 62210 Morelos, M\'exico
\\
             \email{gloria@astro.unam.mx}
             }

  \date{Received January 22, 2019; accepted March 18, 2019}

  \abstract
   {The variability of the Wolf-Rayet star EZ CMa has been documented for close to half a century, and a clear periodicity of $\sim$3.7 days is established. However, all attempts to prove that it is a binary have failed because the photometric, spectroscopic, and polarimetric variations are not coherent over more than a few orbital cycles.}
   {In this letter we show that the lack of coherence in the variability can be explained with a very rapid  apsidal motion in a binary orbit.}
   {We measured the times of minima in a recently published exceptionally long photometric light curve obtained by the Toronto {\emph{BRITE}} satellite. The apsidal motion and the system eccentricity are determined from the length of the time intervals between these minima, which alternate in their duration, following a pattern 
   that is clearly associated with apsidal motion. These minima are superposed on brightness enhancements of the emission from a shock zone, which occur at about the times of periastron phases.}
   {We determine the orbital periodicity, $P_{a}=3.63\, $d, and the period of the apsidal motion, $U\simeq 100\, $d, which together yield an average sidereal period of $P_{s}=3.77\,$d. The eccentricity is found to be close to  0.1. The rate of periapsis retreat changes significantly over the period of observation and is determined to be $-16^\circ\,\mathrm{P}^{-1}_a$ at the beginning of the observing period and $-10^\circ\,\mathrm{P}^{-1}_a$ at the end. 
   }
   {We demonstrate that by introducing a fast apsidal motion, the basic photometric variability is very well explained. The binary nature of EZ CMa is now established. This might imply that other  apparently single Wolf-Rayet stars that emit hard X-rays, similar to EZ CMa, are also binaries.}


\maketitle

\section{Introduction}

Classical Wolf-Rayet (WR) stars are evolved massive objects that are in advanced nuclear burning stages, have shed their outer hydrogen-rich layers, and possess strong stellar winds with enhanced abundances of nuclear processed elements \citep{2000ARA&A..38..143M,2007ARA&A..45..177C, 2012ARA&A..50..107L}.  Two basic channels are believed to give rise to these WRs. The first is the binary interaction channel, whereby the originally more massive star fills its Roche lobe as it leaves the main sequence and transfers mass to its companion, thereby shedding its outer layers and exposing the regions close to the nuclear burning core, as described by \citet{1976IAUS...73...35V}. The second channel is the single-star channel, whereby the outer layers are shed through a combination of stellar wind and violent mass outflow events occurring near the end of the core hydrogen-burning phase \citep{Conti1975}.
The second channel is generally favored for stars lacking  eclipses  or a periodic radial velocity variation curve, even though several of these are extremely variable and have long been suspected to be in binary systems,  the most vexing example of which is EZ CMa.

EZ CMa \citep[HD\,50896 = WR6,][]{vanderHucht2001} is the brightest WR star 
that can be observed from the northern hemisphere.  The variability in its emission lines and light intensity were first reported by \citet{1948PASP...60..383W} and \citet{Ross1961}, respectively.  Subsequent studies disclosed possible variability periods of 1\,d \citep{1967PASP...79...57K,1975A&A....44..219L} and 13\,d \citep{1974PASP...86..767S}, but clear periodicity eluded observers until \citet{1978MmSAI..49..453F} discovered the now well-established 3.76\ d period, confirming it shortly thereafter \citep{Firmani_etal1980}. The authors concluded that EZ CMa is a binary system containing a $\sim$1.3 M$_\odot$ companion, which, they proposed, could be a collapsed companion such as a neutron star because such WR + cc systems are predicted within the scenario proposed by  \citet{1976IAUS...73...35V}.

The collapsed-companion scenario was first challenged by \citet{1988MNRAS.234..783S} and 
\citet{1989ApJ...347..409P}, who argued that the X-ray luminosity is too low for the expected accretion rate onto a neutron star, unless some mechanism is present that inhibits accretion. However, \citet{Skinner_etal2002} analyzed a scenario in which the companion is a non-degenerate low-mass star and showed that the hard component of the observed X-rays could be explained by the shock that is formed when the WR wind collides with the surface of such a companion.  

A  significantly more serious challenge for the binary scenario is the loss of coherence in the 3.76\,d periodic variations when observations covering timescales longer than $\text{about two}$ weeks are analyzed.  This has prompted scenarios invoking a rotating inhomogeneous wind or disk \citep{1991ApJ...368..588U, 1991ApJ...382..301S}  or one with two oppositely directed outflows of enhanced density \citep{1991JRASC..85..214M}, which eventually morphed into models involving corrotating interaction regions, CIRs \citep{1998ApJ...498..413M, St-Louis2018,Moffat2018}.  According to these scenarios, the 3.76\,d period corresponds to the rotation of the WR core, with a yet to be explained instability producing the streams or inhomogeneities in the wind that cause the observed variability.  One common feature of these purported instabilities is  that their location drifts over time with respect to the underlying rotating stellar surface, hence leading to the incoherence when observed over long timescales.

In this paper we solve the puzzle of the non-coherence in the variability of EZ CMa by showing that it is caused by an exceptionally fast apsidal motion of the eccentric binary orbit.  This result is made possible through the recent observations  obtained  by the {\emph{BRITE}} satellite constellation  \citep{Moffat2018}, which provides a nearly continuous data train over 136 days.

\section{Interpretation of the photometric variability}

%
   \begin{figure*}
   \centering
   \includegraphics[width=18cm]{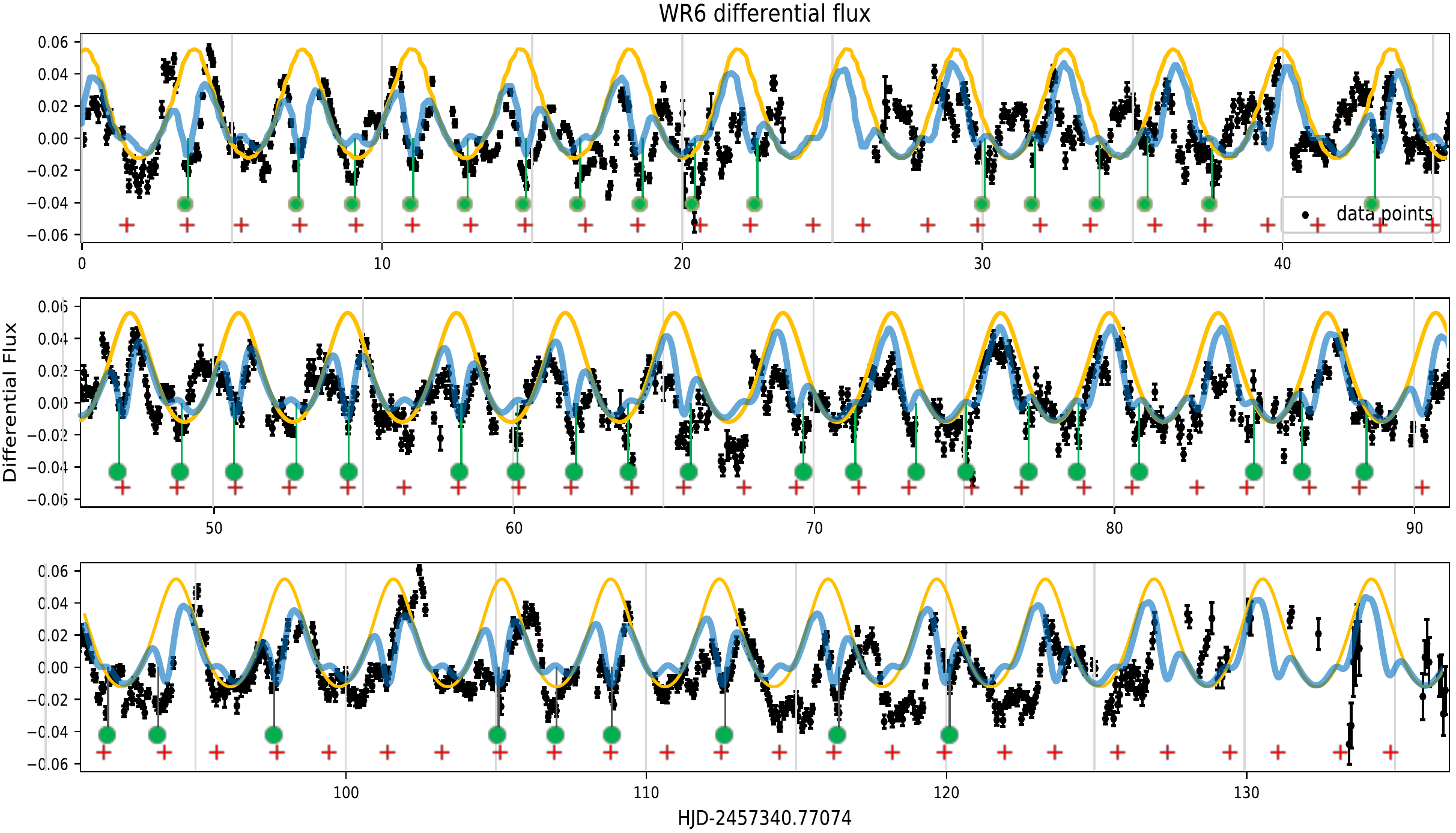}
   \caption{Copy of Fig.\,1 published by \citet{Moffat2018}, displaying the photometric light curve of EZ CMa as observed by the Toronto {\emph{BRITE}} satellite from 2015 November 14 until 2016 March 29. The measured times of well-defined minima are indicated by green dots. The calculated times of upper and lower conjunctions of the orbit, with the parameters given in Table~\ref{Tab2} that are described in the text, are marked by plus signs. The uneclipsed photometric light curve from a shocked zone, which is assumed to vary proportionally to the square of the orbital separation, is given by the orange curve. The light from the shock zone is eclipsed twice each orbit, and the light curve resulting from a simple Gauss-shaped eclipse model is shown by the blue line.
   }
              \label{Fig1}%
    \end{figure*}
\citet{Moffat2018} observed EZ~CMa with the {\emph{BRITE}} satellite constellation from 2015 November 15 to 2016 March 29. We reproduce their published data in our Figure\ \ref{Fig1}.
Adopting the model in which  EZ~CMa is a binary system with a WR primary and a low-mass companion as proposed by \citet{Skinner_etal2002}, we interpret this light curve as an overall $\sim$3.6\,d variation that is twice eclipsed, yielding basically two minima per orbit. The overall variation is due to brightening of the shocked WR wind during the time of periastron passage.  The eclipses are produced when either the WR or the companion occult a portion of the shocked wind emission.   Below we present clear orbital signatures by analyzing the timings of brightness minima.

Light from a shocked wind zone has been identified by \citet{Lamberts_etal2017}, who resolved the WR binary $\gamma^2$~Vel and estimated that the brightness contribution from the shock zone amounts to about 10\,\% of the system's total brightness. \citet{ Richardson_etal2017} showed that the periodic photometric brightening  is produced by spectral line emission and that it varies proportional to the distance between the stars.
Similarly for EZ CMa, we modeled the basic variability as the brightness contribution of the shocked WR wind, which varies with the distance between the objects. Below we derive the eccentricity of the system to be about 0.1, which implies that the emission from the shocked region is variable by $\pm 20$\,\%, if it varies with the square of the separation. The observed variability in the {\emph{BRITE}} satellite filter is on the order of $\pm 0.03$\,mag, which means that for EZ CMa, the shock zone contributes approximately 15\,\% to the total light. The computed uneclipsed light  curve is shown by the orange curve in Figure\ \ref{Fig1}.

The observed light curve then means that the light from the brightest part, that is,\ the most variable part, is eclipsed twice each orbit by either the WR wind or by the as yet undetected companion. Thus, there is a minimum twice in each orbit, similarly as in close binary systems, at both the upper and lower conjunction. We find that basically only the excess emission 
is occulted because the eclipse depth typically reaches the minimum value observed out of eclipse.
As a consequence, the eclipses are most clearly seen when the conjunctions coincide with periastron passages, and less so in the other orbital configurations.  

\section{Determining the apsidal motion}
\begin{figure}
   \centering
   \includegraphics[width=8.5cm]{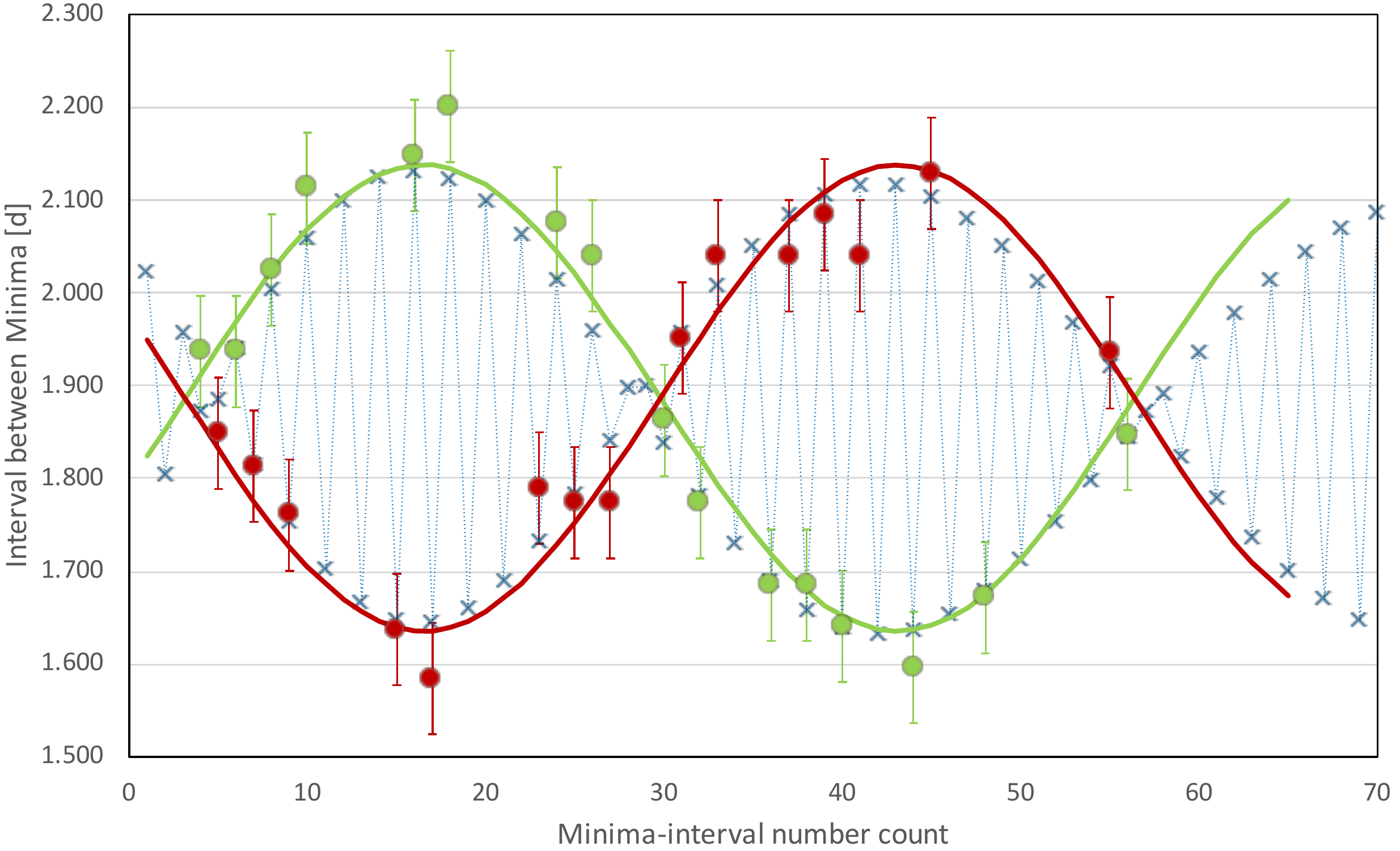}
      \caption{Measured intervals between consecutive minima. The red dots mark the differences in even number minus odd number minima, and green dots show the differences in odd minus even minima. The green and red lines are sine curve fits to the measurements. The blue crosses connected by the blue dotted line show the calculated intervals from the final orbital solution (see text).              }
         \label{Fig2}
   \end{figure}

In the {\emph{BRITE}} light curve we measured the timings of clearly visible minima, which are marked by green dots in Figure\ \ref{Fig1}. We note that the intervals between the minima are not constant but always{\em } alternate systematically between short and long duration, as shown in Figure\ \ref{Fig2}. 
This interval pattern is a distinctive signature of an apsidal motion in an eccentric binary orbit.
  Random drifting of the minima timings on timescales of fractions of the period have been observed in low-mass contact binaries due to reasons other than an apsidal motion \citep{Tran_etal2013}. However, in the case of EZ CMa, we observe that the argument of periastron systematically drifts over more than $360^\circ $, which challenges such explanations.
  Additional evidence for an apsidal motion comes from radial velocities that do not follow the anomalistic phase but agree with the sidereal phase \citep{KSF19}.

%

   \begin{table}
      \caption[]{Orbital parameters derived from a double sine curve fit to the observed minima intervals (Figure\ \ref{Fig2}).}
         \label{Tab1}
     $$ 
         \begin{array}{p{0.5\linewidth}l}
            \hline
            \noalign{\smallskip}
            Eccentricity $\epsilon$ & 0.10\pm 0.01\,  \\ 
            Half sidereal period $\frac{1}{2} P_\mathrm{s}$ & 
            1.88\, \mathrm{d} \\
            Apsidal period $U$  & 100 \pm 5 \, \mathrm{d}  \\
            Time $T^U_0$ of $\omega = 90^{\circ}$ 
            (or $270^{\circ}$)
            & \mathrm{HJD}\,2'457'345.8 \pm 5\, \mathrm{d} \\
            \noalign{\smallskip}
            \hline
         \end{array}
     $$ 
   \end{table}
A double sine curve fit yields the orbital parameters given in Table \ref{Tab1}. The position of the periapsis for the time $T$ is given by the equation
   \begin{equation}
      \omega = 90^\circ + \dot{\omega}\, (T-T^U_0)/P_\mathrm{a}\,,
              \label{Eq1}%
   \end{equation}
where $\dot{\omega}$ is the rate of periapsis advance, $T^U_0$ is the epoch when the periapsis angle is $90^\circ$, and $P_\mathrm{a}$ is the anomalistic period, the time between two passages at periastron. The apsidal motion completes one 360$^\circ$ cycle within the apsidal period $U$. The mean sidereal period, $P_\mathrm{s}$, is the mean time that elapses between consecutive conjunctions, upper or lower, respectively. As there are two eclipses for each orbit, there is half the mean sidereal period equal to the mean interval value. Half the sidereal period and the apsidal period result directly from the sine curve fit, whereas the anomalistic period needs to be calculated from the relation between the three periods,
   \begin{equation}
      P_\mathrm{s} = P_\mathrm{a}\, 
      (1-\dot{\omega}/360^\circ)\,,
              \label{Eq2}%
   \end{equation}
 where the apsidal motion in degrees per anomalistic period is given by $\dot{\omega}=360^\circ P_\mathrm{a} / U$.
 
When 
$\dot{\omega}$ is negative, which is the case for EZ CMa, the subtraction in Eq.\,2 changes to an addition in the relation
    \begin{equation}
      P_\mathrm{s} = P_\mathrm{a}\, 
      (1+P_\mathrm{a}/U)\,,
              \label{Eq3}%
   \end{equation}
if the apsidal period $U$ is given as a positive number. The fact that the periastron angle is retreating yields the sidereal period $P_\mathrm{s}=3.77\,$d, which is longer than the anomalistic period, $P_\mathrm{a}=3.64$\,d. This can be verified by counting in Fig.\,\ref{Fig1} that there is one more light maximum than eclipse-pairs over the 100 days of the apsidal period $U$. 

\section{Interpretation of the {\em O-C} diagram}
%
   \begin{figure}
   \centering
   \includegraphics[width=8.5cm]{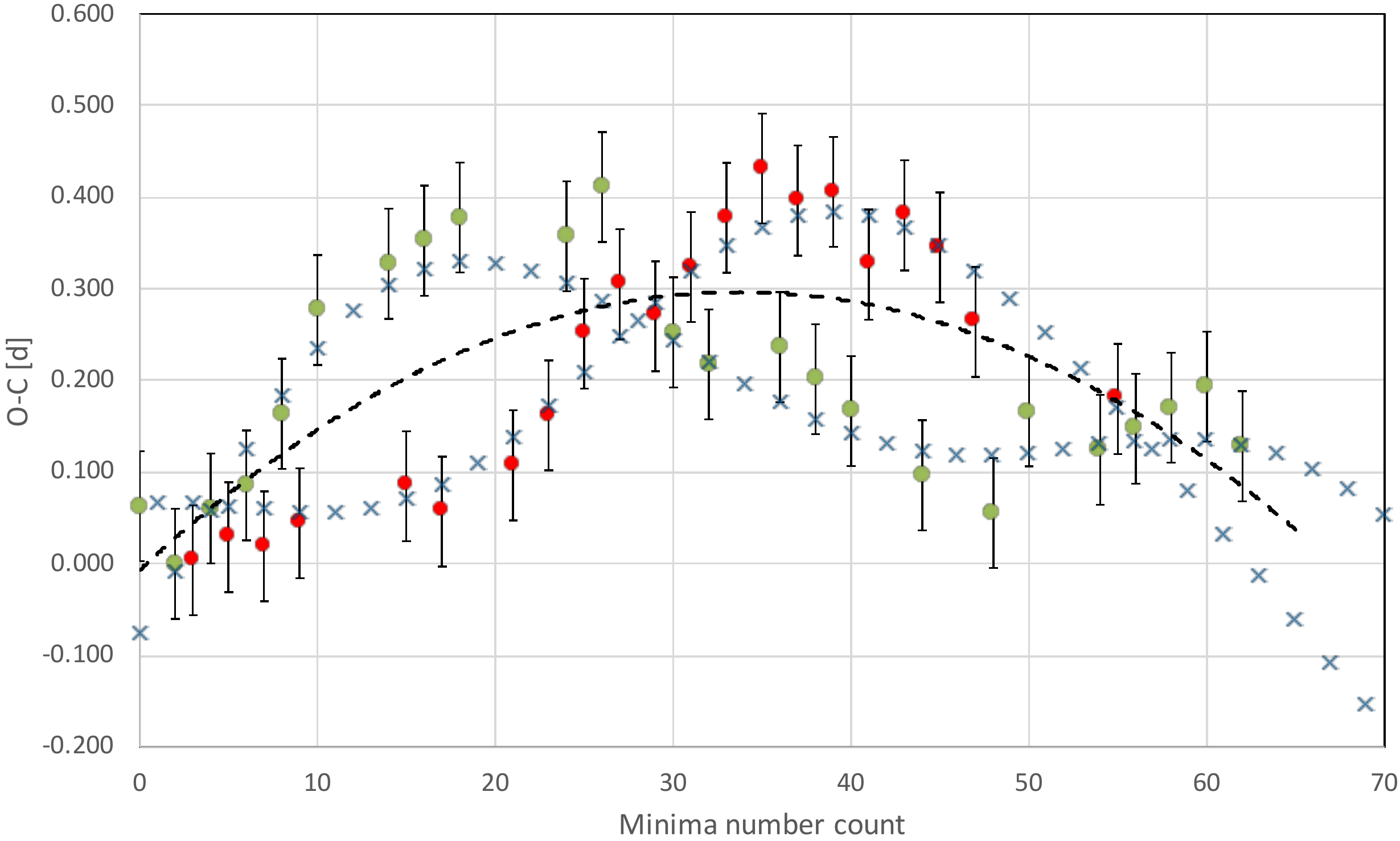}
      \caption{Observed epochs of minima, marked with red dots for even number minima and with green dots for odd minima, minus computed minima epochs calculated with the relation $T(E)=E_0+P\,E$, with $P=1.88$\,d, which is half the sidereal period. $E$ is the number count of the minima starting with zero. The dashed curve is a quadratic fit to all measured minima. The blue crosses mark the calculated minima from the final orbital solution relative to the linear ephemeris.
      }
         \label{Fig3}
   \end{figure}
For a clear detection of an apsidal motion we have used the intervals between successive minima (Fig.\ \ref{Fig2}). For the purpose of determining an apsidal motion, a {\em O-C} diagram may be used as well, which illustrates the difference between an observed minimum epoch and one computed with a linear ephemeris. In Figure\ \ref{Fig3} we display the {\em O-C} diagram for the measured minima, which are indicated in Figure\ \ref{Fig1}. The apsidal motion is also clearly visible in the {\em O-C} diagram, but in addition, an overall parabolic trend is indicated by the quadratic fit. This dependence indicates a changing period. From the measured minima timings, it cannot be determined which period is changing. The {\em O-C} diagram can be fitted equally well with a change in sidereal period as with a change of the anomalistic period. \\

If the changing period were the anomalistic period, then the observed trend seen in Figure\ \ref{Fig3} would require $\dot{P_\mathrm{a}} = -5.7\,10^{-4}\,$d/d. This rate is  too fast given the reported periods for EZ\ CMa in the literature, which are all of about 3.7\,d  \citep[see][]{Georgiev_etal1999}.  An estimate of a possible period decrease, by including historic data, yields a much lower value,
$\dot{P_\mathrm{a}}
\simeq -1\,10^{-5}\,\mathrm{d/d}$, which is a considerably fast period change but not sufficient to reproduce the observed shape in the {\em O-C} diagram.\\

We therefore fit the {\em O-C} measurements of Figure\ \ref{Fig3} allowing for a change in apsidal motion. 
Equation~1 is now modified to read
\begin{equation}
      \omega = 90^\circ + \dot{\omega}\, (T-T^U_0)/P_\mathrm{a}
      + \frac{1}{2}\,\ddot{\omega}\, 
      \left( 
      \frac{(T-T^U_0)}{P_\mathrm{a}}
      \right)^2\,,
              \label{Eq4} %
\end{equation}
where $\ddot{\omega}$ accounts for the change in rate of the periapsis retreat. 

We solved the binary motion for the full time interval of Figure\ \ref{Fig1} in time steps of 0.1 days with the changing argument of periastron as given by Eq.~4. From this resulting table of projected positions in the sky, $(x(t), y(t), z(t))$, we determined the epochs of conjunction by searching for the sign change in the projected x-coordinate and interpolated between the two epochs of the adjacent grid points.
The fit comprises six parameters, which were used to minimize  the difference squared between the calculated minima epochs and the measurements as well as the difference between the calculated and observed intervals shown in Figur\,\,\ref{Fig2}. Table\ \ref{Tab2} lists the best-fit orbital parameters.
   \begin{table}
      \caption[]{Orbital parameters derived from fitting the measurements of minima epoches shown in Figs.\ \ref{Fig3} and simultaneously the intervals shown in Figure\ \ref{Fig2}. The uncertainties are estimated by fitting only one set of measurements, as shown either in Fig.\ \ref{Fig2} or in Fig.\  \ref{Fig3}.}
         \label{Tab2}
     $$ 
         \begin{array}{p{0.5\linewidth}l}
            \hline
            \noalign{\smallskip}
            Eccentricity $\epsilon$ &   0.102\pm 0.01\, \\
            Anomalistic period $P_\mathrm{a}$ & 3.626\pm 0.01\, \mathrm{d}\\
            Time of periapsis $E_0$ & 
            \mathrm{HJD}\,2'457'348.1\ \pm 0.3\, \mathrm{d} \\
            Apsidal motion 
            $\dot{\omega}(T^U_0)$  & -15.6^\circ \pm 1^\circ \,\mathrm{ P}^{-1}_a   \\
            Apsidal motion change $\ddot{\omega}$ & 
            0.17^\circ \pm 0.06^\circ\,\mathrm{P}^{-2}_a \\
            Time $T^U_0$\ of $\omega = 90^{\circ}$ 
            (or $270^{\circ}$) &
            \mathrm{HJD}\,2'457'350.3 \pm 1\, \mathrm{d} \\
            \noalign{\smallskip}
            \hline
         \end{array}
     $$ 
   \end{table}

\section{Discussion}
 We have derived the orbital parameters exclusively from the timing of the epochs of mid-eclipse, and did not use the shape of the observed light curve. We now construct a synthetic light curve based on our eclipse model by subtracting a Gaussian function at each eclipse from the shocked wind light curve.  
 The free parameters are the depth and the width of the Gaussian, and the inclination of the system.
 In Figure\ \ref{Fig1} we have chosen the inclination $i=74^\circ$ and adjusted the depth. Visually, however, the same result is obtained with $i=60^\circ$ and a larger depth. 
We find that basically the two eclipses are identical, that is,\ when 
the shock zone is eclipsed by either the WR or the unseen object. The resulting light curve is plotted in Figure\ \ref{Fig1} with the blue line. The agreement with the observations is surprisingly good given the simplicity of 
  our eclipse and wind collision models.

From the results of the hydrodynamic simulations made for the $\gamma^2$~Vel system, it is clear that the reality of eclipsing the shock zone is much more complex than can be represented by a simple Gaussian curve. It is also clear that the emission from the shock zone, which basically
consists of emission lines, is variable \citep{Richardson_etal2017, Lamberts_etal2017}. 
Therefore, the calculated light curve may be shifted up or down, depending on whether a part of the zone is in a higher or lower state. Such a misfit usually persists for several orbital periods at the same phase.  It is interesting to note that the two segments of data for which the simple model has the most difficulty in reproducing the minima occur $\sim$100 days apart; or in other words, at the same orbital geometry as defined by $\omega$.

We adopted for the mass of the WR star $M_\mathrm{WR}\simeq 20\,\mathrm{M}_\odot$, determined from the helium-star mass-luminosity relation \citep{SM92}, where the luminosity was derived independently of distance from the energy needed to accelerate the WR wind  \citep{Schmutz1997}\footnote{The spectroscopic analyses of EZ CMa published by
\citet{Schmutz1997} adopted a distance of $d\simeq 1.8$\,pc as determined by \citet{HS1995}. 
ESA's \emph{GAIA} satellite measures $\pi=0.41\pm 0.05\,$mas, yielding $d\simeq 2.4\,$pc, and based on this, the system is about 20 to 50\% more distant than assumed in his analysis. 
However, allowing for 15 to 20\% of the light to come for the shock zone, and allowing another 10\%\ for other objects in the system, we conclude that the correction needed for a larger distance is cancelled and that \citet{Schmutz1997} assumed the correct order of magnitude for the absolute brightness of the WR component. Thus, the stellar parameters of the WR star given by \citet{Schmutz1997} can be considered valid estimates. This holds in particular for the electron-scattering radius of the WR wind that is used in the discussion section.}.
For the companion we may follow the proposal of \citet{Skinner_etal2002} that the companion is a  low-mass star (LMS),
and we adopt $M_\mathrm{LMS}\approx 1.5\,\mathrm{M}_\odot$. 
The combined mass and the anomalistc period yield a separation of the two objects of $a\simeq 0.13\,$AU.  
This value can be compared with the optically thick (electron scattering) radius of the WR star given in Figure\ 3 of \citet{Schmutz1997}, $R_\mathrm{WR}\approx 5\,\mathrm{R}_*$, where $R_*=3.5\,\mathrm{R}_\odot$, which yields $R_\mathrm{WR}\approx 0.08\,$AU, or $0.6\,a$. 
This ratio compares favorably to the value of the width of the Gauss curve,  $w\approx 0.6\mathrm{\ to\ }0.65\,a$, which is needed to describe the eclipse light curve in Figure\ \ref{Fig1}.

The eclipse by the unseen object has about the same duration as that by the WR star. This is possible if the eclipsed part of the shock zone, that is, the brightest variable part, is very close to the other object such that the object covers about a quarter of the sky seen from the shock. This would be indeed the case if the other object were a low-mass star, as speculated above. The shock zone is a direct result from the collision of the wind with 
this star, or, somewhat speculatively, with an accretion disk or magnetopause.

In order to determine the orbital phases and orientation of the orbit during observations obtained in the past (e.g., those of XMM-Newton from 2001), a much more precise set of orbital and precession parameters is needed, which requires more observations.  Moreover, the very fast change rate that we find in the apsidal motion, $\ddot\omega$, cannot be sustained for very long.
This also requires further investigation, as does the physical mechanism that could produce the fast apsidal motion. The primary assumption is that it is caused by a third body. A rough estimate of the minimum period of the outer body, using the equations given by \citet{Borkovits_etal2015}, neglecting tidal and relativistic effects, yields $P_\mathrm{out} \approx 22$\,d if $m_3 \simeq m_\mathrm{WR}$ or 31\,d if $ m_3/m_{123}\approx 1$, where m$_3$ and m$_{123}$ are the mass of the third body and the total system mass,
respectively \citetext{personal communication by the referee}. Such a short period may be revealed by searching for a variable inner systemic velocity in the radial velocity curves of the WR star. 
Further investigation is also needed to refine the wind collision model that provides the periastron times and to search for the signatures of the associated emission lines in optical spectra.

\section{Conclusions\label{Conclusion}}

The long-duration light curve of 136 days allowed us to identify the minima epochs and to detect a regular pattern that is characteristic for an apsidal motion. The timing measurements of the minima were used to determine six parameters of the orbit of the system.
Based on these parameters, and using
a simple Gauss-shaped eclipse model and a simple wind-collision model producing enhanced emission at periastron, we calculated a light curve that reproduces the observed complicated light curve shape over the full observational time-span. 
No doubt remains that EZ\ CMa is a binary.

Many eccentric binaries are known to exhibit an apsidal motion 
\citep{Wolf2006, 2008MNRAS.388.1836W}, 
which in extreme cases is dominated by the perturbations of a third body.  The high value of $\dot{\omega}$ and the fact that instead of an advancing periapsis it is receding
are very unusual, however. In principle, the presence of a third body could also explain the curvature in the {\em O-C} diagram by the light travel time in the outer orbit. If this were the case, then the radial velocity variations of the inner system would have to be greater than $100$\,km/s, and we expect that this would have been noted in previous investigations.

Our findings have implications for the determination of the stellar parameters of the WR star because an analysis has to take into account that there is light from a shock zone and possibly also from other objects in the system.
The binary scenario also changes the view of \citet{Morel_etal1997} and \citet{St-Louis2018}. What they interpreted as additional emission from corotating interaction regions are the brightening episodes of the shock zone. As illustrated in Figure~\ref{Fig1}, the large majority of the observed light maxima are explained by the corresponding binary phase.

On a more general level, we note that the interpretation of the observed X-ray spectrum of EZ CMa by \citet{Skinner_etal2002} has lead to the correct conclusion. They argued that the presence of hard X-rays implies the presence of a companion. When we extend this reasoning to other apparently single nitrogen-type WR stars with X-ray detections, then we find that all of them show an admixture of cool ($kT_1 < 1$\, keV) and hot ($kT_2 > 2$\, keV) X-ray plasma and prominent emission lines from ions such as Si\,{\sc xiii} and S\,{\sc xv} \citep{Skinner_etal2010}. This indicates that they might have as yet undetected companions. The most extreme candidate after EZ CMa is WR\,134. 
This in turn would mean that the binary scenario for forming WR stars is more important, if not dominant, and that current evolution models of massive single stars assume too large an amount of mass is lost by single massive stars in order to force the simulations to reach a WR stage.

\begin{acknowledgements}
We thank an anonymous referee for very constructive and helpful comments. Based on data collected by the {\em BRITE}
Constellation satellite mission, designed, built, launched, operated and supported by the Austrian Research Promotion Agency (FFG), the University of Vienna, the Technical University of Graz, the University of Innsbruck, the Canadian Space Agency (CSA), the University of Toronto Institute for Aerospace Studies (UTIAS), the Foundation for Polish Science \& Technology (FNiTP MNiSW), and National Science Centre (NCN).
GK acknowledges support from CONACYT grant 252499.
\end{acknowledgements}

\ 

\ 

\ 
\bibliographystyle{aa} 
\bibliography{35094_references} 

\end{document}